\documentclass[aps,twocolumn,showpacs,prl]{revtex4}

\usepackage{dcolumn}
\usepackage{amsmath}
\usepackage{graphicx}

\begin{document}

\title{Effective attraction induced by repulsive interaction in a spin-transfer system}

\author{Ya.\ B. Bazaliy}
 \affiliation{Instituut Lorentz, Leiden University, The Netherlands,}
 \affiliation{Department of Physics and Astronomy, University of South Carolina, Columbia, SC,}
 \affiliation{Institute of Magnetism, National Academy of Science, Ukraine.}

\date{April, 2006}

\begin{abstract}
In magnetic systems with dominating easy-plane anisotropy the
magnetization can be described by an effective one dimensional
equation for the in-plane angle. Re-deriving this equation in the
presence of spin-transfer torques, we obtain a description that
allows for a more intuitive understanding of spintronic devices'
operation and can serve as a tool for finding new dynamic regimes. A
surprising prediction is obtained for a planar ``spin-flip
transistor'': an unstable equilibrium point can be stabilized by a
current induced torque that further repels the system from that
point. Stabilization by repulsion happens due to the presence of
dissipative environment and requires a Gilbert damping constant that
is large enough to ensure overdamped dynamics at zero current.
\end{abstract}

\pacs{72.25.Pn, 72.25.Mk, 85.75.-d}

\maketitle

%\section{Introduction}
In physics, there are cases where due to the presence of complex
environment a repulsive force can lead to actual attraction of the
entities. A well known example is a superconductor, where the Cooper
pairs are formed from electrons repelled by the Coulomb forces due
to the dynamical elastic environment. Here we report a phenomena of
effective attraction induced by the repulsive spin-transfer torque
in the presence of highly dissipative environment. The spin-transfer
effect producing the repulsive torque is a non-equilibrium
interaction that arises when a current of electrons flows through a
non-collinear magnetic texture \cite{berger,slon96,bjz1998}. This
interaction can become significant in nanoscopic magnets and is
nowadays studied experimentally in a variety of systems. Its
manifestations - either current induced magnetic switching
\cite{layered} or magnetic domain wall motion \cite{dw} - serve as
an underlying mechanism for a number of suggested memory and logic
applications.

Here we consider a conventional spin-transfer device consisting of a
a magnetic polarizer (fixed layer) and a small magnet (free layer)
with electric current flowing from one to another (Fig 1). Both
layers can be described by a macro-spin model due to large exchange
stiffness. The free layer is influenced by the spin transfer torque,
while the polarizer is too large to feel it. Magnetic dynamics of
the free layer is described by the Landau-Lifshitz-Gilbert (LLG)
equation with the spin transfer torque term \cite{slon96,bjz2004}.

\begin{figure}[b]
    \resizebox{.45\textwidth}{!}{\includegraphics{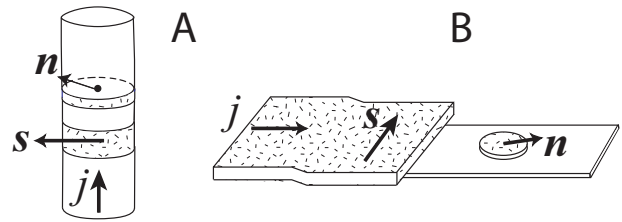}}
\caption{Planar spin-transfer devices. Hashed parts of the devices
are ferromagnetic, white parts are made from a non-magnetic metal.}
 \label{fig:devices}
\end{figure}

The solutions of LLG are easy to find for the simplest easy axis
magnetic anisotropy of the free layer. There exists a critical
current at which the free layer either switches between the two
minima of magnetic energy, or goes into a state of permanent
precession, powered by the current source
\cite{slon96,bjz2004,sun2000}. The same basic processes happen in
the case of realistic anisotropies, however the complexity of the
calculations increases substantially. In a nanopillar device
\cite{katine2000} one additionally finds that stabilization of
magnetic energy maxima is possible (``canted states''
\cite{bjz2004}) and that multiple precession modes exist with
transitions between them happening as the current is increased
\cite{sun2000,kiselev2003,xiao2005}. The anisotropy of a nanopillar
device is a combination of a magnetic easy plane and magnetic easy
axis directed in that plane. Experimentally, the easy plane
anisotropy energy is usually much larger than the easy axis energy,
i.e. the system is in the regime of a planar spintronic device
\cite{bauer-planar-review} (Fig.~\ref{fig:devices}). This limit of
dominating easy plane energy is characterized by another
simplification of the dynamic equations \cite{weinan-e,boj2007},
which comes not from the high symmetry of the problem, but from the
existence of a small parameter: the ratio of the energy modulation
within the plane to the easy plane energy. The deviation of the
magnetization from the plane becomes small, making the motion
effectively one dimensional.

In this paper we present a general form of effective planar equation
describing a macrospin free layer in the presence of spin transfer
torques. Its relationship to the first order expansion in the
current magnitude used in Ref.~\onlinecite{boj2007} is discussed at
the end. We then use this equation to study the ``spin-flip
transistor'': a planar device in which the spin polarizer is
perpendicular to the direction favored by the magnetic anisotropy
energy. It was predicted \cite{bauer-spin-flip-transistor} that the
competition between the anisotropy and spin transfer torques leads
to a 90 degrees jump of the magnetization at the critical current.
Whether the jump happens into the parallel or antiparallel state
with respect to the polarizer is determined by the direction of the
current.

Here it is shown that the behavior of the spin-flip transistor is
more complicated than expected from the simple picture above.
Namely, the current inducing a jump into the parallel direction can
also stabilize the antiparallel direction. This conclusion is
certainly counter-intuitive because the spin torque repels the
magnetization from this already unstable saddle point of the energy.
However, a combination of two destabilizing torques manages to
result in a stable equilibrium. We will see that this happens due to
the dissipation terms and a sufficiently large (but still small
compared to unity) Gilbert damping constant is required to observe
the phenomena.

The magnetization of the free layer ${\bf M} = M {\bf n}$ has a
constant absolute value $M$ and a direction given by a unit vector
${\bf n}(t)$. The LLG equation \cite{slon96,bjz2004} reads:
\begin{equation}
 \label{eq:vector_LLG}
{\dot {\bf n}} = \frac{\gamma}{M} \left[ - \frac{\delta E}{\delta
{\bf n}} \times {\bf n} \right] + u({\bf n}) [{\bf n} \times [{\bf
s} \times {\bf n}]] + \alpha [{\bf n} \times \dot {\bf n}] \ .
\end{equation}
Here $\gamma$ is the gyromagnetic ratio, $E({\bf n})$ is the
magnetic energy of the free layer, and $\alpha$ is the Gilbert
damping constant. The second term on the right is the spin transfer
torque, where ${\bf s}$ is a unit vector along the direction of the
polarizer, and the spin transfer strength $u({\bf n})$ is
proportional to the electric current $I$ \cite{bjz2004,boj2007}. In
general, spin transfer strength is a function of the angle between
the polarizer and the free layer $u({\bf n}) = f[({\bf n}\cdot{\bf
s})] \ I$, with the function $f[({\bf n}\cdot{\bf s})]$ being
material and device specific. Equation (\ref{eq:vector_LLG}) can be
written in polar angles $(\theta(t),\phi(t))$:
\begin{eqnarray}
 \nonumber
\dot\theta + \alpha \dot\phi \sin\theta &=&
    - \frac{\gamma}{M\sin\theta} \frac{\partial E}{\partial\phi}
    + u ({\bf s} \cdot {{\bf e}_{\theta}}) \equiv F_{\theta} \ ,
 \\
 \label{eq:polar_angles_LLG}
\dot\phi \sin\theta - \alpha \dot\theta &=&
    \frac{\gamma}{M} \frac{\partial E}{\partial\theta}
    +  u ({\bf s} \cdot {\bf e}_{\phi}) \equiv F_{\phi}\ ,
\end{eqnarray}
with tangent vectors ${\bf e}_{\phi} = [\hat z \times {\bf
n}]/\sin\theta$, ${\bf e}_{\theta} = [{\bf e}_{\phi} \times {\bf
n}]$. The easy plane is chosen at $\theta = \pi/2$, and the magnetic
energy has the form $E = (K_{\perp}/2)\cos^2\theta +
E_r(\theta,\phi)$, where $E_r$ is the ``residual'' energy. In the
planar limit, $K_{\perp} \to \infty$, the energy minima are very
close to the easy plane and the low energy solutions of LLG have the
property $\theta(t) = \pi/2 + \delta\theta$ with $\delta\theta \to
0$. Equations (\ref{eq:polar_angles_LLG}) can then be expanded in
small parameters $|E_r|/K_{\perp} \ll 1$, $|u({\bf n})|/K_{\perp}
\ll 1$. Assuming time-independent $u$ and $\bf s$ we obtain an
effective equation of the in-plane motion
\begin{equation}
 \label{eq:effective_equation}
\frac{1}{\omega_{\perp}} \ddot\phi +  \alpha_{eff}\dot\phi = -
\frac{\gamma}{M} \frac{\partial E_{eff}}{\partial\phi} \ ,
\end{equation}
which has has the form of the Newton's equation of motion for a
particle in external potential $E_{eff}(\phi)$ with a variable
viscous friction coefficient $\alpha_{eff}(\phi)$. The expressions
for the effective friction and energy  are
\begin{eqnarray}
 \label{eq:effective_alpha}
 && \alpha_{eff}(\phi) = \alpha -
 (\Gamma_{\phi} + \Gamma_{\theta})/\omega_{\perp} \ ,
 \\
 \nonumber
 &&  \Gamma_{\phi} =
      \left( \partial F_{\phi}/\partial\phi \right)_{\theta = \pi/2}
   , \
   \Gamma_{\theta} = \left(\partial F_{\theta}/\partial\theta\right)_{\theta =
   \pi/2} \ ,
\end{eqnarray}
and
\begin{eqnarray}
 \label{eq:effective_energy}
 && E_{eff}(\phi) = E_r\left(\pi/2,\phi \right) + \Delta E(\phi) \ ,
 \\
 \nonumber
 && \Delta E =
 - \frac{M}{\gamma} \int^{\phi}
 \left[ u({\bf n}) ({\bf s}\cdot {\bf e}_{\theta})  -
  \frac{\Gamma_{\theta}}{\omega_{\perp}} F_{\phi} \right]_{\theta = \frac{\pi}{2}}
 \ d\phi' \ .
\end{eqnarray}
Equation (\ref{eq:effective_equation}) with definitions
(\ref{eq:effective_alpha},\ref{eq:effective_energy}) gives a general
description of a planar device in the presence of spin transfer
torque. At non-zero current the effective friction can become
negative (see below), and the effective energy is not necessarily
periodic in $\phi$ (e.g. in the case of ``magnetic fan''
\cite{boj2007,wang2006}). Physically this reflects the possibility
of extracting energy from the current source, and thus developing a
``negative dissipation'' in the system.

In many planar devices the polarizer direction ${\bf s}$ lies in the
easy plane, $\theta_s = \pi/2$, with a direction defined by the
azimuthal angle $\phi_s$. At the same time the residual energy has a
property $(\partial E_r/\partial\theta)_{\theta = \pi/2} = 0$, i.e.
does not shift the energy minima away from the plane. We will also
use the simplest form $f[({\bf n}\cdot{\bf s})] = {\rm const}$ for
the spin transfer strength. A more realistic function will not
change the result qualitatively and can be easily used if needed.
With these restrictions the effective friction and the energy
correction get the form:
\begin{eqnarray}
\label{eq:alpha_DeltaE_special}
 \alpha_{eff} &=& \alpha +
 \frac{2 u \cos(\phi_s - \phi)}{\omega_{\perp}}
 \\
 \nonumber
 \Delta E &=&
 - \frac{M u^2}{2\gamma\omega_{\perp}}\cos^2(\phi_s - \phi)  \ .
\end{eqnarray}

%\section{Analysis of the spin-flip transistor}
In a spin-flip transistor the polarizer direction is given by
$\phi_s = \pi/2$. Following
Ref.~\onlinecite{bauer-spin-flip-transistor}, we consider in-plane
anisotropy energy $E_r(\pi/2,\phi) = - (K_{||}/2)\cos^2\phi$
corresponding to an easy axis. Then the effective friction is
$\alpha_{eff} = \alpha + (2 u\sin\phi)/\omega_{\perp}$ and effective
energy equals $(\gamma/M)E_{eff} = -[(\omega_{||} -
u^2/\omega_{\perp})/2]\cos^2\phi  \, + \, {\rm const}$ with
$\omega_{||} = \gamma K_{||}/M$. Equilibrium points $\phi = 0,\pm
\pi/2,\pi$ are the minima and maxima of the effective energy, and do
not depend on $u$. Stability of any equilibrium in one dimension
depends on whether it is a minimum or a maximum of $E_{eff}$ and on
the sign of $\alpha_{eff}$ at the equilibrium point. It is easy to
check, that out of four possibilities only an energy minimum with
$\alpha_{eff} > 0$ is stable. In the case of a spin-flip transistor
the energy landscape changes above a threshold $|u|
> \sqrt{\omega_{||}\omega_{\perp}}$: the energy minima at $\phi =
0,\pi$ become maxima, and, vice versa, the energy maxima at $\phi =
\pm\pi/2$ switch to minima. Effective friction at $\phi = 0,\pi$ is
positive independent of $u$, while at $\phi = \pm\pi/2$ it changes
sign at $u = \mp \alpha\omega_{\perp}/2$.

\begin{figure}[t]
    \resizebox{.45\textwidth}{!}
    {\includegraphics{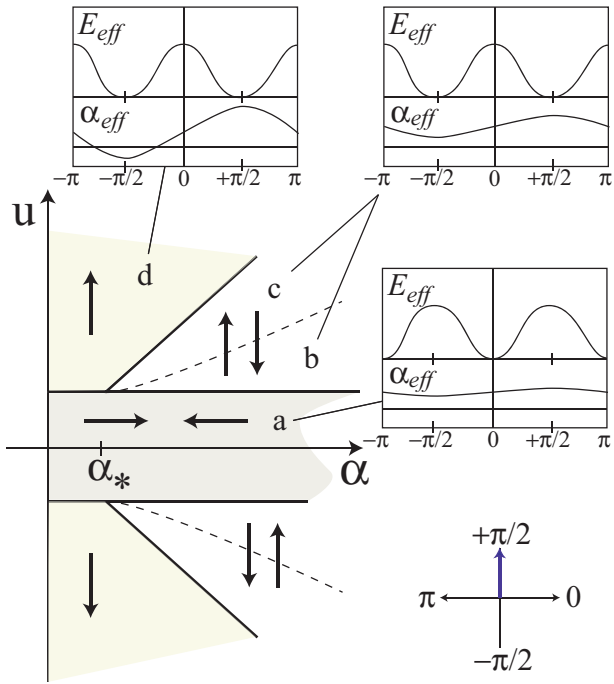}}
\caption{Switching diagram of the spin-flip transistor. In each zone
one or two arrows show the possible stable directions of the free
layer magnetization. Directions of the easy axis and spin polarizer
are defined in the right bottom corner. Angular dependencies of
$\alpha_{eff}$ and $E_{eff}$ are given in insets. Stable subregions
``b'' and ``c'' differ in overdamped vs. underdamped approach to the
equilibrium.}
     \label{fig:switching_diagram}
\end{figure}

The behavior of the spin-flip transistor is summarized in a
switching diagram Fig.~\ref{fig:switching_diagram} plotted on the
plane of the material characteristic $\alpha$ and the experimental
parameter $u \sim I$. For definiteness we will discuss a current
with $u>0$. The effect of the opposite current is completely
symmetric. For small values of Gilbert damping one observes
stabilization of the $\phi = \pi/2$ (parallel) equilibrium to which
the spin torque attracts the magnetization of the free layer, while
the opposite (antiparallel) direction remains unstable. This is in
accord with the results of
Ref.~\onlinecite{bauer-spin-flip-transistor}. However, when the
damping constant is larger than the critical value $\alpha_{*} =
2\sqrt{\omega_{||}/\omega_{\perp}}$, a window of stability of the
antiparallel equilibrium opens on the diagram. Since $\alpha \ll 1$,
a sufficiently large easy plane energy is required to achieve
$\alpha_{*} < \alpha \ll 1$.

\begin{figure}[t]
    \resizebox{.49\textwidth}{!}{\includegraphics{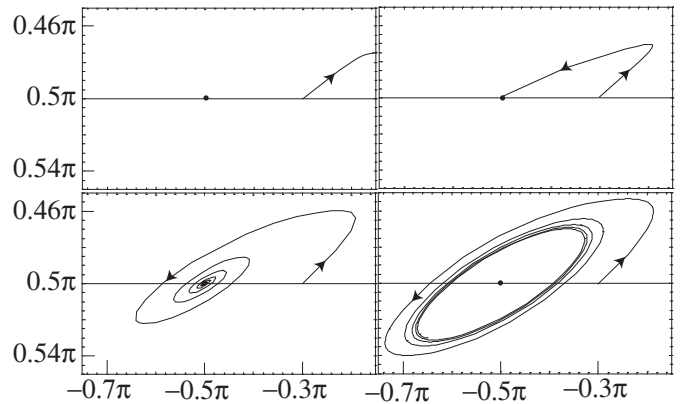}}
\caption{Typical trajectories of ${\bf n}(t)$ for
$\omega_{||}/\omega_{\perp} = 0.01$, $\alpha = 1.5 \, \alpha_{*}$.
The plot labels correspond to the regions in
Fig.~\ref{fig:switching_diagram}, the current magnitude is given in
the units of $u/\sqrt{\omega_{||}/\omega_{\perp}}$ and we look at
the stability of the $\phi = -\pi/2$ equilibrium: (a) 0.93, unstable
(b): 1.08, stabilized with overdamped approach (c): 1.38, stable,
but with oscillatory approach (d): 1.53, unstable; a stable cycle is
formed around the equilibrium.}
 \label{fig:trajectories}
\end{figure}

If one thinks about the stability of the $(\theta,\phi) = (\pi/2,
-\pi/2)$ equilibrium for $u>0$ in terms of
Eq.~(\ref{eq:vector_LLG}), this prediction seems completely
unexpected. The anisotropy torques do not stabilize this equilibrium
because it is a saddle point of the total magnetic energy $E$, and
the added spin transfer torque repels ${\bf n}$ from this point as
well. The whole phenomena may be called ``stabilization by
repulsion''. To check the accuracy of the planar approximation
(\ref{eq:effective_equation}), the result was verified using the LLG
equations (\ref{eq:polar_angles_LLG}) with no approximations for the
axis-and-plane energy $E = (K_{\perp}/2)\cos^2\theta -
(K_{||}/2)\sin^2\theta\cos^2\phi$. Calculating the eigenvalues of
the linearized dynamic matrices \cite{bjz2004} at the  equilibrium
points $(\pi/2, \pm\pi/2)$ we obtained the same switching diagram
and confirmed the stabilization of the antiparallel direction.
Typical trajectories ${\bf n}(t)$ numerically calculated from the
LLG equation with no approximations are shown in
Fig.~\ref{fig:trajectories} to illustrate the predictions.
%of the effective planar description.
At $u > \sqrt{\omega_{||}\omega_{\perp}}$ the $\phi = -\pi/2$
equilibrium is stabilized. In accord with the predictions of
Eqs.~(\ref{eq:effective_equation}),(\ref{eq:alpha_DeltaE_special}),
the wedge of its stability consists of two regions (b) and (c)
characterized by overdamped and underdamped dynamics during the
approach to the equilibrium. The dividing dashed line is given by $u
= \omega_{||}/\alpha + \alpha\omega_{\perp}/4$. It was checked that
small deviations of the polarizer $\bf s$ from the $(\pi/2,\pi/2)$
direction do not change the behavior qualitatively. Larger
deviations eventually destroy the effect, especially the
out-of-plane deviation which produces the ``magnetic fan'' effect
\cite{wang2006} leading to the full-circle rotation of $\phi$ in the
plane.

As the current is further increased to $u > \alpha\omega_{\perp}/2$,
the antiparallel state looses stability and the trajectory
approaches a stable precession cycle
(Fig.~\ref{fig:trajectories}(d)). The existence of the precession
state is easy to understand from (\ref{eq:effective_equation})
viewed as an equation for a particle in external potential. Just
above the stability boundary the effective friction
$\alpha_{eff}(\phi)$ is negative in a small vicinity of $\phi =
-\pi/2$, and positive elsewhere. Within the $\alpha_{eff}<0$ region
the dissipation is negative and any small deviation from the
equilibrium initiates growing oscillations. As their amplitude
exceeds the size of that region, part of the cycle starts to happen
with positive dissipation. Eventually the amplitude reaches a value
at which the energy gain during the motion in the $\alpha_{eff} < 0$
region is exactly compensated by the energy loss in the
$\alpha_{eff} > 0$ region: thus a cycle solution emerges. The
effective planar description allows for the analysis of the further
evolution of the cycle with transitions into different precession
modes, which will be a subject of another publication.

The fact that $\alpha > \alpha_*$ condition is required for the
stabilization means that dissipation terms play a crucial role
entangling two types of repulsion to produce a net attraction to the
reversed direction. Note that an interplay of a strong easy plane
anisotropy and dissipation terms produces unexpected effects already
in conventional ($u=0$) magnetic systems. The effective planar
equation (\ref{eq:effective_equation}) at $u=0$ was discussed in
Ref.~\onlinecite{weinan-e}. It was found that the same threshold
$\alpha_{*}$ represents a boundary between the oscillatory and
overdamped approaches the equilibrium. Above $\alpha_{*}$ the
familiar precession of a magnetic moment in the anisotropy field is
replaced by the dissipative motion directed towards the energy
minimum. When the easy plane anisotropy is strong enough to ensure
$\alpha \gg \alpha_{*}$, one can drop the second order time
derivative term in Eq.~(\ref{eq:effective_equation}) and use the
resulting first order dissipative equation. In the presence of spin
transfer, $\alpha_{eff}(\phi,u)$ depends on the current and can
assume small values even for $\alpha \gg \alpha_{*}$, thus no
general statement about the $\ddot\phi$ term can be made.

The simplest easy axis energy expression $E_r(\pi/2,\phi) = -
(K_{||}/2)\cos^2\phi$ happens to have the same angular dependence as
$\Delta E(\phi)$ given by Eq.~(\ref{eq:alpha_DeltaE_special}). Due
to this special property the energy profile flips upside down at $u
= \sqrt{\omega_{||}\omega_{\perp}}$. For a generic $E_r(\pi/2,\phi)$
with minima at $\phi = 0,\pi$ and maxima at $\phi = \pm\pi/2$ the
nature of equilibria will change at different current thresholds.
This will make the switching diagram more complicated, but will not
affect the stabilization by repulsion phenomena. Similar
complications will be introduced by a generic $f[({\bf n}\cdot{\bf
s})]$ angular dependence of the spin transfer strength.

In Ref.~\onlinecite{boj2007} the known switching diagram for the
collinear ($\phi_s = 0$) devices \cite{bjz2004,kiselev2003,xiao2005}
were reproduced by equation (\ref{eq:effective_equation}) with
$E_{eff} = E_r(\pi/2,\phi)$. The $\Delta E$ term
(\ref{eq:alpha_DeltaE_special}) was dropped as being second order in
small $u$. This approximation gives a correct result for the
following reason. In a collinear device $(\gamma/M)E_{eff} =
-[(\omega_{||} + u^2/\omega_{\perp})/2]\cos^2\phi  + {\rm const}$
and the current never changes the nature of the equilibrium from a
maximum to a minimum. Consequently, dropping $\Delta E$ does not
affect the results. As was already noted in
Ref.~\onlinecite{boj2007}, the first order expansion in $u$ is
insufficient for the description of a spin-flip transistor, where
the full form (\ref{eq:alpha_DeltaE_special}) is required.

%\section{Conclusions}
In summary, we derived a general form of the effective planar
equation (\ref{eq:effective_equation}) for a macrospin free layer in
the presence of spin transfer torque produced by a fixed
spin-polarizer and time-independent current. Qualitative
understanding of the solutions of planar equation is obtained by
employing the analogy with a one-dimensional mechanical motion of a
particle with variable friction coefficient in an external
potential. The resulting predictive power is illustrated by the
discovery of the stabilization by repulsion phenomena in the
spin-flip device. Such stabilization relies on the form of the
dissipative torques in the LLG equation and happens only for a large
enough Gilbert damping constant. The new stable state and the
corresponding precession cycle can be used to engineer novel memory
or logic devices, and microwave nano-generators with tunable
frequency.

To observe the phenomena experimentally, one has to fabricate a
device with $\alpha > \alpha_{*}$, and initially set it into a
parallel or antiparallel state by external magnetic field. Then the
current is turned on and the field is switched off. Both states
should be stabilized by a moderate current
$\sqrt{\omega_{||}\omega_{\perp}} < u < \alpha\omega_{\perp}/2$, but
cannot yet be distinguished by their magnetoresistive signals. The
difference can be observed as the current is increased above the
$\alpha\omega_{\perp}/2$ threshold: the parallel state will remain a
stable equilibrium, while the antiparallel state will transform into
a precession cycle and an oscillating component of magnetoresistance
will appear.

%\section{Acknowledgments}
The author wishes to thank C. W. J. Beenakker, G. E. W. Bauer, and
Yu. V. Nazarov for illuminating discussions. Research at Leiden
University was supported by the Dutch Science Foundation NWO/FOM.
Part of this work was performed at KITP Santa Barbara supported by
the NSF grant No. PHY99-07949, and at Aspen Physics Institute during
the Winter program of 2007.


\begin{thebibliography}{2}

\bibitem{berger} L. Berger, J. Appl. Phys., {\bf 49}, 2160 (1978);
Phys. Rev. B  {\bf 33}, 1572 (1986);
%``Possible existence of a Josephson effect in ferromagnets''
% deflection of the domain wall and precession onset for j>j_c
J. Appl. Phys. {\bf 63}, 1663 (1988).
%``Exchange interaction between electric current and magnetic domain wall
% containing Bloch lines''
% (Bloch lines are moved along the domain wall due to the current)

\bibitem{slon96} J. Slonczewski, J. Magn. Magn. Mater. {\bf 159}, L1
(1996).

\bibitem{bjz1998} Ya. B. Bazaliy {\em et al.},
%, B. A. Jones, and Shou-Cheng Zhang,
Phys. Rev. B, {\bf 57}, R3213 (1998).
%Modification of the Landau-Lifshitz equation in the presence of a spin-polarized current in
%colossal- and giant-magnetoresistive materials

\bibitem{layered}
 S. Kaka {\em et al.},
 Nature {\bf 437}, 389 (2005);
%Mutual phase-locking of microwave spin torque nano-oscillators
 M. R. Pufall {\em et al.},
 Phys.Rev. Lett. {\bf 97}, 087206 (2006);
%Electrical measurement of spin-wave interactions of proximate spin transfer nanooscillators
 M. L. Schneider {\rm et al.},
 Appl. Phys. Lett., {\bf 90}, 092504 (2007);
%Thermal effects on the critical current of spin torque switching in spin valve nanopillars
 X. Jiang {\em et al.},
 Phys. Rev. Lett. {\bf 97}, 217202 (2006);
%Temperature dependence of current-induced magnetization switching in spin valves
%with a ferrimagnetic CoGd free layer
 W. Chen {\em et al.},
 Phys. Rev. B, {\bf 74}, 144408(2006);
%Spin transfer in bilayer magnetic nanopillars at high fields as a
%function of free-layer thickness
 B. Ozyilmaz {\em et al.},
 Phys. Rev. Lett., {\bf 93}, 176604 (2004);
%  Current-induced excitations in single cobalt ferromagnetic layer nanopillars
 I. N. Krivorotov {\em et al.}
 Science, {\bf 307}, 228 (2005);
%Time-domain measurements of nanomagnet dynamics driven by
%spin-transfer torques
 N. C. Emley {\em et al.}
 Phys. Rev. Lett., {\bf 96}, 247204 (2006);
%Time-resolved spin-torque switching and enhanced damping in
%Permalloy/Cu/Permalloy spin-valve nanopillars
 J. C. Sankey, {\em et al.},
 Phys. Rev. Lett., {\bf 96}, 227601 (2006).
%Spin-transfer-driven ferromagnetic resonance of individual nanomagnets

\bibitem{dw}
 G. Beach {\em et al.},
 Phys. Rev. Lett., {\bf 97}, 057203 (2006);
%Nonlinear domain-wall velocity enhancement by spin-polarized electric current
 Nature Materials, {\bf 4}, 741 (2005);
%Dynamics of field-driven domain-wall propagation in ferromagnetic nanowires
 M. Klaui {\em et al.},
 Phys. Rev. Lett., {\bf 95}, 026601 (2005);
%Direct observation of domain-wall configurations transformed by spin currents
 M. Laufenberg {\em et al.},
 Phys. Rev. Lett., {\bf 97}, 046602 (2006);
%Temperature dependence of the spin torque effect in current-induced
%domain wall motion
 L. Thomas {\em et al.},
 Science, {\bf 315}, 1553 (2007);
%Resonant amplification of magnetic domain-wall motion by a train of
%current pulses
 M. Hayashi {\em et al.},
 Phys. Rev. Lett., {\bf 98}, 037204 (2007);
%Current driven domain wall velocities exceeding the spin angular
%momentum transfer rate in permalloy nanowires
 Nature Physics, {\bf 3}, 21 (2007);
%Direct observation of the coherent precession of magnetic domain walls
%propagating along permalloy nanowires
  Phys. Rev. Lett., {\bf 97}, 207205 (2006);
%Dependence of current and field driven depinning of domain walls on
%their structure and chirality in permalloy nanowires
 M. Yamanouchi {\em et al.}
 Nature, {\bf 428}, 539 (2004);
%Current-induced domain-wall switching in a ferromagnetic semiconductor structure
 Phys. Rev. Lett., {\bf 96}, 096601 (2006).
 %Velocity of Domain-Wall Motion Induced by Electrical Current
 %in the Ferromagnetic Semiconductor (Ga,Mn)As

\bibitem{bjz2004} Ya. B. Bazaliy {\em et al.},
Phys. Rev. B, {\bf 69}, 094421 (2004).
%``Current-induced magnetization switching in small domains of different anisotropies"

\bibitem{sun2000}
J. Z. Sun, Phys. Rev. B {\bf 62}, 570 (2000).

\bibitem{katine2000}
 J. A. Katine  {\em et al.},
 Phys. Rev. Lett., {\bf 84}, 3149 (2000).
%Current-Driven Magnetization Reversal and Spin-Wave Excitations in Co/Cu/Co Pillars

\bibitem{kiselev2003}
S. I. Kiselev  {\em et al.}, Nature, {\bf 425}, 380 (2003).
%Microwave oscillations of a nanomagnet driven by a spin-polarized current

\bibitem{xiao2005}
J. Xiao  {\em et al.}, Phys. Rev. B, {\bf 72}, 014446 (2005)
%Macrospin models of spin transfer dynamics

\bibitem{bauer-planar-review}
A. Brataas {\em et al.},
%G. E. W. Bauer, and P. J. Kelly,
Phys. Rep., {\bf 427}, 157 (2006).
%[cond-mat/0602151].

\bibitem{weinan-e} C. Garcia-Cervera {\em et al.}, J. Appl. Phys.,
{\bf 90}, 370 (2001).
%``Effective dynamics for ferromagnetic thin films"

\bibitem{boj2007} Ya. B. Bazaliy  {\em et al.},
arXiv:0705.0406v1 (2007), to be published in J. Nanoscience and
Nanotechnology.

\bibitem{bauer-spin-flip-transistor}
A. Brataas {\em et al.}, Phys. Rev. Lett. {\bf 84}, 2481 (2000);
%
X.~Wang  {\em et al.}, Japan. J. Appl. Phys., {\bf 45}, 3863 (2006).
%, cond-mat/0601632 (2006).
%``Current-controlled magnetization dynamics in the spin-flip transistor''

\bibitem{wang2006}
X. Wang {\em et al.}, Phys. Rev. B, {\bf 73}, 054436 (2006).
% ``magnetic fan''

\end{thebibliography}
\end{document}